%% file: main.tex
\documentclass[conference,table]{IEEEtran}

\IEEEoverridecommandlockouts

\usepackage{hyperref}
\usepackage{bbm}
\usepackage{nicefrac}
\usepackage[switch]{lineno}
\usepackage[utf8]{inputenc}
\usepackage{nicematrix}
\usepackage[T1]{fontenc}
\usepackage{mathptmx}
\usepackage{etoolbox}
\usepackage{amsmath,amsfonts}
\usepackage{algorithmic}
\usepackage{textcomp}
\usepackage{booktabs}
\usepackage{comment}
\usepackage{float}
\usepackage[final]{pdfpages}
\usepackage{tablefootnote}
\usepackage{caption}
\usepackage{subcaption}
\usepackage{amsmath}
\usepackage{flushend}
\usepackage{balance}
\usepackage{ulem}
\usepackage[bottom]{footmisc}
\usepackage{tabularx}
\usepackage{colortbl}
\usepackage{xargs}
\usepackage[colorinlistoftodos,prependcaption]{todonotes}

\setuptodonotes{inline,color=blue!30}
\newcommandx{\unsure}[2][1=]{\todo[linecolor=red,backgroundcolor=red!25,bordercolor=red,#1]{#2}}
\newcommandx{\change}[2][1=]{\todo[linecolor=blue,backgroundcolor=blue!25,bordercolor=blue,#1]{#2}}
\newcommandx{\info}[2][1=]{\todo[linecolor=OliveGreen,backgroundcolor=OliveGreen!25,bordercolor=OliveGreen,#1]{#2}}
\newcommandx{\improvement}[2][1=]{\todo[linecolor=Plum,backgroundcolor=Plum!25,bordercolor=Plum,#1]{#2}}
\newcommandx{\thiswillnotshow}[2][1=]{\todo[disable,#1]{#2}}

\definecolor{sam}{RGB}{28, 28, 28}
\definecolor{zhen}{RGB}{0, 143, 67}

\usepackage{multirow}
\usepackage{balance}

\begin{document}

\title{A Comprehensive Performance Study of Large Language Models on Novel AI Accelerators}


\makeatletter
\newcommand{\linebreakand}{%
  \end{@IEEEauthorhalign}
  \hfill\mbox{}\par
  \mbox{}\hfill\begin{@IEEEauthorhalign}
}
\makeatother


\author{\IEEEauthorblockN{Murali Emani\IEEEauthorrefmark{1}}
  \IEEEauthorblockA{memani@anl.gov}
  \and
  \IEEEauthorblockN{Sam Foreman\IEEEauthorrefmark{1}}
  \IEEEauthorblockA{foremans@anl.gov}
  \and
   \IEEEauthorblockN{Varuni Sastry\IEEEauthorrefmark{1}}
  \IEEEauthorblockA{vsastry@anl.gov} 
  \and
  \IEEEauthorblockN{Zhen Xie\IEEEauthorrefmark{2}}
  \IEEEauthorblockA{zxie3@binghamton.edu}
  \and
  \IEEEauthorblockN{Siddhisanket Raskar\IEEEauthorrefmark{1}}
  \IEEEauthorblockA{sraskar@anl.gov}
  \and
  \IEEEauthorblockN{William Arnold\IEEEauthorrefmark{1}}
  \IEEEauthorblockA{arnoldw@anl.gov}
    \linebreakand 

  \and
  \IEEEauthorblockN{Rajeev Thakur\IEEEauthorrefmark{1}}
  \IEEEauthorblockA{thakur@anl.gov}
  \and
 \IEEEauthorblockN{Venkatram Vishwanath\IEEEauthorrefmark{1}}
  \IEEEauthorblockA{venkat@anl.gov}
  \and
  \IEEEauthorblockN{Michael E. Papka\IEEEauthorrefmark{1}\IEEEauthorrefmark{3}}
  \IEEEauthorblockA{papka@anl.gov}
  \linebreakand
  
\IEEEauthorrefmark{1}Argonne National Laboratory, Lemont, IL 60439, USA, \\
\IEEEauthorrefmark{2}State University of New York, Binghamton, NY, 13092, USA, \\
\IEEEauthorrefmark{3}University of Illinois, Chicago, IL 60637, USA 
}

\maketitle


\begin{abstract}

Artificial intelligence (AI) methods have become critical in scientific applications to help accelerate scientific discovery. Large language models (LLMs) are being considered as a promising approach to address some of the challenging problems because of their superior generalization capabilities across domains. The effectiveness of the models and the accuracy of the applications is contingent upon their efficient execution on the underlying hardware infrastructure. Specialized AI accelerator hardware systems have recently become available for accelerating AI applications. However, the comparative performance of these AI accelerators on large language models has not been previously studied. In this paper, we systematically study LLMs on multiple AI accelerators and GPUs and evaluate their performance characteristics for these models. We evaluate these systems with (i) a micro-benchmark using a core transformer block, (ii) a GPT- 2 model, and (iii) an LLM-driven science use case, GenSLM. We present our findings and analyses of the models' performance to better understand the intrinsic capabilities of AI accelerators.
Furthermore, our analysis takes into account key factors such as sequence lengths, scaling behavior, sparsity, and sensitivity to gradient accumulation steps.
\end{abstract}

\pagestyle{plain}
\thispagestyle{plain}
\input{tex/01_Introduction}

\input{tex/03_LLM}
\input{tex/04_Evaluation}
\input{tex/05_Results}
\input{tex/06_Conclusion}

\newpage
\bibliographystyle{IEEEtran}
\bibliography{references}

\end{document}

%% file: tex/01_Introduction.tex
\section{Introduction}

The incorporation of Artificial Intelligence (AI) for Science has gained increasing interest in scientific research institutes and supercomputing facilities, with the goal of accelerating scientific discovery with novel approaches involving AI. This synergy has increased interest in the adoption of novel AI-driven techniques to help develop solutions to complex scientific problems such as protein-structure prediction \cite{alphafold, alphafold2}, cosmology parameter prediction \cite{cosmoflow}, neutrino particle detection \cite{cosmictagger}, drug design for precision medicine \cite{candle}, genome-scale foundation model \cite{genslm} and weather forecasting model \cite{fourcastnet}. 
Some of the most commonly used AI techniques include convolutional neural networks, recurrent neural networks, graph neural networks, and large language models (LLMs). These techniques, with their unique architectural characteristics, have become invaluable to assist scientists in their research.
Within the AI landscape, the domain of Natural language processing (NLP) has experienced a massive surge in growth. fostering the development of LLMs to use in various tasks such as question answering, text summarization, and language translation. These models are becoming increasingly critical in scientific machine learning applications. 

\begin{table*}[htbp]
\begin{center}
    \caption{Features of evaluated AI accelerators}     \label{table:hwoverview}
\footnotesize
\centering
    \begin{NiceTabular}{ p{3.0cm} | p{2cm} | p{2cm} | p{2cm} | p{2cm} |p{2cm} | p{2cm}  }  
    \CodeBefore
        \rowcolor{gray!20}{1}
        \rowcolors{2}{gray!15}{white}
    \Body
        \toprule
        \footnotesize{\textbf{Feature}} & \footnotesize{\textbf{Nvidia A100}} & \footnotesize{\textbf{SambaNova DataScale SN30}}  & \footnotesize{\textbf{Cerebras CS-2}} & \footnotesize{\textbf{Graphcore BowPod64}} & 
        \footnotesize{\textbf{Habana Gaudi2}} & \footnotesize{\textbf{AMD MI250}}\\
        \toprule
        System Size\footnotemark[1] & $64\left(=16\times 4\right)$& 
        $64\left(= 8 \times 8\right)$& 
        $2 \left(= 1 \times 2\right)$&
        $64 \left(= 4 \times 16 \right)$&
        $8\left(= 1 \times 8 \right)$& 
        $4\left(= 1 \times 4 \right)$\\
        
        Memory \footnotesize{($/ \mathrm{node}$)} & 
        160 GB &
        8 TB & 
        1 TB &
        3.6 GB/128 GB \footnotemark[2]&
        768 GB &
        512 GB \\

        Memory \footnotesize{($/ \mathrm{device}$)} & 
        40 GB &
        1 TB & 
        1 TB &
        900 MB/32 GB &
        96 GB &
        128 GB \\

        Interconnect & 
        NVLink & 
        Ethernet-based & 
        Ethernet-based & 
        IPU Link & 
        RoCE\footnotemark[3] &
        AMD CDNA \\    
        
        Software Stack\footnotemark[4] & 
        \texttt{TF}, \texttt{PT}, 	\texttt{ONNX}, \texttt{MxNET}, \texttt{CUDA} &
        \texttt{SambaFlow\textsuperscript{TM}}, \texttt{PT}, \texttt{TF} &
        \texttt{PT}, \texttt{Cerebras SDK} &
        \texttt{TF}, \texttt{PT}, \texttt{ONNX}, \texttt{PopArt} &
        \texttt{Synapse AI}, \texttt{TF} and \texttt{PT} &
        \texttt{TF}, \texttt{PT}, \texttt{ROCm} \\

        Precision\footnotemark[5] & 
        \texttt{TF32, FP32, FP16, BF16} & 
        \texttt{FP32, BF16, Int32, Int16, Int8} & 
        \texttt{FP32, FP16, BF16, cbfloat} & 
        \texttt{FP32, FP16} & 
        
        \texttt{FP32, TF32, BF16, FP16, FP8} &
        \texttt{FP64, FP32, FP16, BF16, INT8, INT4} \\
        
        Compute Units \footnotesize{($/ \mathrm{device}$)}& 
        6912 Cuda Cores, 432 Tensor Cores & 
         1280 PCUs & 
        850,000 Cores & 
        1472 Compute cores &	
         
        24 TPC + 2 MME

         & 
        13312 cores, 208 compute units  \\

      \bottomrule
    \end{NiceTabular}
\end{center}
\end{table*}

LLMs, such as Generative Pre-trained Transformers (GPT) GPT-3 \cite{gpt3}, 
LLaMA \cite{llama}, Llama 2 \cite{llama2}, and Bloom \cite{bloom} have seen a massive improvement in their complexity along with the quality of results for these tasks.
This growth has been driven in part by the rapid emergence of transformer-based models as the \textit{de-facto} architecture for both traditional applications and a potent tool for scientific use cases. Transformer-based architectures have found a multitude of applications, from accelerating drug discovery to understanding genetic sequences. 
%
%
%
For example, GenSLM \cite{genslm} provides an LLM-based foundation model to predict Sars-CoV2 variants of concern. Its strength lies in its capacity to inform the design of effective anti-viral drugs.
The GensLM model is trained on an extensive dataset of over 110 million raw nucleotide sequences and with model scales between 25 million to 25 billion trainable parameters.
%
%
However, training GPT-variant LLMs with large model parameters and longer sequence lengths necessitates specialized computing resources and innovative implementation techniques and optimizations in the software stack.

To address these requirements, AI accelerators, designed on non-traditional architectures, such as dataflow, have emerged. 
These accelerators are custom-built to efficiently support AI workloads with their powerful hardware compute engines and novel software optimizations. They are proven to effectively train several AI models, with a special focus on LLMs. 
%
%
%
With their unique system characteristics, these AI accelerators are empowered to tackle the challenges posed by LLMs. 
Besides their capability in training, these accelerators are able to run some of the largest GPT models, besides providing a suite of pre-trained GPT models \cite{sn-llm-1, sn-gpt-suite, cs-gpt-suite, dey2023cerebrasgpt, gc-gpt, gc-model-garden, habana-gpt-1, habana-gpt-2}. These models demonstrate the versatility and scalability of AI accelerators.  
With the increase in the size and complexity of LLMs, innovative training techniques have become critical, such as sparsity \cite{sambanova-sparsity, spdf, saxena2023sparse, sparsityv100}, 
to further enhance the training of LLMs with billions of parameters.

Evaluation of LLMs on diverse hardware platforms is of crucial importance to understand the capabilities and limitations of both traditional and non-traditional architectures.
Prior work has studied LLMs on leadership class supercomputers \cite{llmonfrontier} and with traditional deep learning benchmarks \cite{emani2022benchmarking, YIN2021100005} providing detailed insights into their capabilities. However, no comprehensive evaluation has been performed across a variety of AI accelerators, especially for LLMs.
This paper aims to address this gap with a detailed performance evaluation of language models on multiple AI accelerators and is the first of its kind benchmarking study to our best knowledge.
The major contributions of this paper are:
\begin{itemize}
    \item A systematic evaluation of LLMs on state-of-the-art AI accelerators. 
    \item Focus on a transformer block micro-benchmark that is a core component in GPT-based models.
    \item Comprehensive evaluation of GPT-2 XL 1.5B parameter model to glean insights into model performance across all systems.
    \item Porting and evaluation of a science application: GenSLM, a foundation model for gene sequencing.
    \item Studying the impact of sequence lengths, sparsity, and gradient accumulation steps on model throughput.
\end{itemize}

We present an overview of LLMs and the various AI accelerators in Section \ref{section:overview}, followed by details of the evaluated models, namely the transformer block micro-benchmark, GPT-2 XL, and GenSLM application in Section \ref{section:evaluation}. We describe in Section \ref{section:accelerators} the implementation of the LLMs on different AI accelerators. We present experimental results  in Section \ref{section:results}, followed by conclusions in Section \ref{section:conclusion}.

%% file: tex/03_LLM.tex
\section{Overview of Large Language Models and AI Accelerators} 
\label{section:overview}



\footnotetext[1]{Note: \texttt{System Size} is reported as $N_{\texttt{devices}}\left(= N_{\texttt{nodes}} \times \frac{\texttt{devices}}{\texttt{node}}\right)$.}

\footnotetext[2] {Reported as SRAM/DRAM for device and node memory}

\footnotetext[3]{RDMA over Converged Ethernet}
\footnotetext[4]{\texttt{TF}: TensorFlow, \texttt{PT}: PyTorch.}
\footnotetext[5]{We include the most commonly used floating precision data types.}

{Large language models are a type of artificial intelligence system that use deep learning algorithms to process and generate natural language text. These models have become increasingly popular in recent years due to their ability to perform a wide range of language-related tasks, such as machine translation, text summarization, and question answering. The development of large language models has been driven by advances in deep learning techniques, particularly in the area of transformer models. These models use self-attention mechanisms to process text input, allowing them to capture complex patterns and relationships within language data. They are also pretrained on large datasets using unsupervised learning techniques, such as masked language modeling and next-sentence prediction, which help them to learn a broad range of language features and structures. Fine-tuning on specific tasks further improves their performance and adaptability.}

{One of the most well-known LLMs is the GPT (Generative Pretrained Transformer) series, which was developed by OpenAI and is used to } answer questions, translate languages, and generate text.
These tasks are realized by generative pre-training of a language model
on a diverse data corpus of unlabeled text, followed by discriminative fine-tuning on a 
specific task. Task-aware input
transformations during fine-tuning help achieve effective transfer with 
minimal changes to the model architecture. 
Owing to the neural architectural differences, GPT models can broadly be classified into GPT \cite{gpt}
GPT-2 \cite{gpt-2}, GPT-3 \cite{gpt3}, and more recently GPT-4~\cite{gpt4}.

\setcounter{footnote}{5}

AI accelerators comprised of GPU-based systems and novel non-traditional AI hardware, such as dataflow architectures, have proven to boost the efficiency of diverse AI models. Below we describe the accelerators used in this study, while the configurations are listed in Table \ref{table:hwoverview}.

\textbf{\textit{Nvidia A100:}}{
The A100 GPU consists of 6912 CUDA cores and 432 Tensor cores to accelerate parallel workloads.
On a single DGX node, there are 4 A100 GPUs, interconnected with NVLink.
We use a forked 
implementation of Microsoft's \texttt{Megatron-DeepSpeed}\footnote{Which is itself a fork of NVIDIA's original \href{https://github.com/nvidia/Megatron-LM}{Megatron-LM}~\cite{Shoeybi2019MegatronLMTM}} framework for our evaluations.
In doing so, we can take full advantage of DeepSpeed's various optimizations and convenient features such as ZeRO offloading and automatic metric tracking (with communication + FLOP profiling).
}
All experiments were conducted on the Polaris~\cite{polaris} supercomputer at the Argonne Leadership Computing Facility (ALCF) with 4 A100 GPUs and 40 GB memory per node.

\textbf{\textit{Cerebras CS-2:}} 
The Cerebras CS-2 is a wafer-scale deep learning accelerator comprising 850,000 processing cores, each providing 48KB of dedicated SRAM memory for an on-chip total of 40 GB and interconnected to optimize bandwidth and latency.
The system has been scaled to two CS-2 wafer-scale engines nodes interconnected by the SwarmX fabric and with the MemoryX memory subsystem to enable large models. The wafer-scale cluster supports a weight streaming execution mode \cite{WSmode}  where each model layer is loaded one by one.  This feature enables users to run large language models in which each layer’s weights fit in memory, but not the entire model.
The software platform integrates popular machine learning frameworks such as PyTorch.
With a single Cerebras CS-2, the largest model supported is the GPT-3 (175B parameter model) and the CS-2 can support sequence lengths up to 58k.

\textbf{\textit{SambaNova SN30:}} The SambaNova DataScale system uses the second-generation Reconfigurable Dataflow Unit (RDU) processor for optimal dataflow processing and acceleration. 
Each RDU has 1280 Pattern Compute Units (PCU) and 1 TB of off-chip memory.
The system consists of eight nodes, each with eight RDUs interconnected to enable model and data parallelism.
SambaFlow, its software stack, extracts, optimizes and maps dataflow graphs to the RDUs from the PyTorch machine learning framework.
SN30 can train models up to 176 B parameters 
on 4 RDUs. 

\textbf{\textit{Graphcore Bow Pod64}}: The Graphcore 22 petaflops Bow Pod64 system is the latest next-generation accelerator from Graphcore.
It is a one-rack system consisting of 64 Bow-class IPUs with a custom interconnect.
The Graphcore software stack includes the Poplar SDK and has support for TensorFlow and PyTorch.
The Bow system currently supports GPT-3 175B parameter model with 256 IPUs.

\textbf{\textit{Habana Gaudi2:}} The Habana Gaudi2 processor features two Matrix Multiplication Engines (MME), 24 fully programmable VLIW SIMD tensor processor cores, integrating 24 100 GbE ports of RDMA over Converged Ethernet (RoCE) into each processor chip to efficiently scale training. The Gaudi system consists of one HLS-2 server with eight Gaudi2 HL-225H cards. The software stack comprises the SynapseAI stack and supports TensorFlow and PyTorch. It supports the existing deep learning optimization library DeepSpeed and the customized library Optimum Habana, which is the interface between the transformer library and Habana’s Gaudi processor (HPU). On the Gaudi system, the largest model currently validated is the GPT-3 (175B parameter model) running on 384 Gaudi2 cards.

\textbf{\textit{AMD MI250:}} The AMD MI250 GPU is based on CDNA2 architecture and consists of 13,312 stream processors distributed across 208 compute units coupled with 128 GB of dedicated HBM2e memory with 3.276 TB/s of memory bandwidth.
It is able to achieve 362.1 TFlops of FP16 and 45.3 TFlops of FP32 peak performance.
Each GPU, comprising of two tiles, is connected to the host using PCIe Gen4 and uses InfiniBand for to inter-node communication.
AMD ROCm open software platform supports the common DL stack, including Tensorflow and PyTorch, in addition to libraries such as rocBLAS, rocSPARSE, rocFFT, and RCCL (ROCm Collective Communication Library).

%% file: tex/04_Evaluation.tex
\section{Evaluation}
\label{section:evaluation}

In this work, we primarily focused on evaluating (i) transformer benchmarks, (ii) the GPT 2-XL model, and (iii) a science application, GenSLM \cite{genslm}, a foundation model for genome sequencing.

(1) \textbf{Transformer micro-benchmark}:
{To evaluate the performance of a transformer benchmark on AI accelerators, several key factors must be considered. First, it is important to choose appropriate micro-benchmarks that reflect the workload of the transformer models being used. Once the suitable micro-benchmarks have been selected, it is necessary to collect performance metrics, such as throughput, which can be done by measuring the number of input tokens processed per second to complete a certain number of iterations. Additionally, it is important to monitor the utilization of hardware resources, such as memory and compute units. Finally, it is recommended to compare the performance of traditional NVIDIA GPUs against other AI accelerators to gain a more comprehensive understanding of their strengths and weaknesses. By carefully evaluating these factors, it is possible to effectively predict the performance of transformer models on AI accelerators.}

\begin{figure}[htpb]
\centering
 \vspace{-0.1in}
 \includegraphics[width=0.5\textwidth]{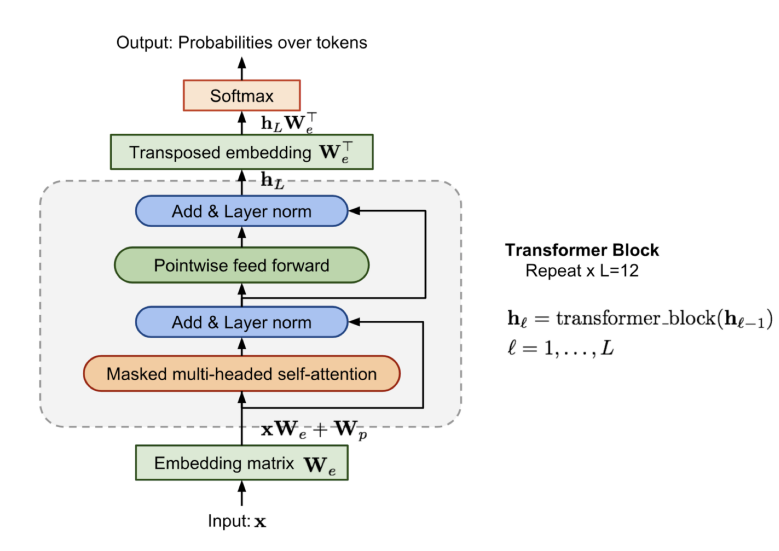}
 \caption{GPT transformer block~\cite{vaswani2017attention}}
 \label{fig:gpt-block}
\end{figure}

{
A \textit{transformer block} (Figure~\ref{fig:gpt-block}) is a widely recognized and established micro-benchmark for transformer models. The transformer block is an ideal choice for a micro-benchmark due to several reasons. First, it is a widely used building block in natural language processing tasks such as language modeling and text generation, making it a relevant benchmark for many LLM applications. Second, a transformer block is relatively simple and small in size compared to larger transformer models, which makes it easier to run and evaluate. This also allows for faster experimentation in evaluating new hardware architectures. Finally, the transformer block includes many of the key components of transformer models, including self-attention and feedforward layers, making it a suitable representative of transformer models in general. Overall, the transformer block is a well-established and widely accepted micro-benchmark, making it an excellent choice for evaluating the performance of LLM models.}

{A transformer block consists of a multi-head self-attention layer followed by a feedforward neural network layer. The input to the transformer block is a sequence of tokens. In this work, the length of the input sequence we evaluated is 1024. To calculate the FLOPs of the transformer block, we need to consider the number of operations required for each layer. The self-attention layer requires $O$($n^2d$) FLOPs, where $n$ is the sequence length and $d$ is the hidden dimension. The feedforward neural network layer requires $O$($ndk$) FLOPs, where $k$ is the size of the hidden layer. Therefore, the total number of FLOPs for the transformer block is $O$($n^2d$ + $ndk$).}

(2) \textbf{GPT-2 XL}:
For this study, we used the GPT-2 XL 1.5B parameter model for 
{pre-training} experimentation to analyze how performant the accelerators are in running large language models. Though the evaluated systems can individually support much larger models and varied configurations as stated in Section~\ref{section:overview}, 
we chose GPT-2 XL because it can be easily implemented on each of the systems in a timely manner for a fair comparison. Also, the memory and compute requirements of a GPT-2-sized model fit well with the minimum unit of compute node on each of the systems; hence the insights gained here can be extended to help drive the decisions in choosing accelerators that can yield optimal performance for any given large transformer-based model architecture. The dataset used in this evaluation is Open Web Text (OWT) \cite{owt} which is an open-source version of the WebText corpus. It consists of 38 GB of text data from 8,013,769 documents, which constitute content extracted from URLs shared on Reddit with at least three upvotes. 
For this model, we measured the model throughput across a scale of devices on each system. Additionally, we evaluated the impact of sequence lengths, gradient accumulation steps (GAS), and sparsity on the model performance.

\begin{figure}[htpb]
    \centering
    \vspace{-1em}
    \includegraphics[width=\linewidth]{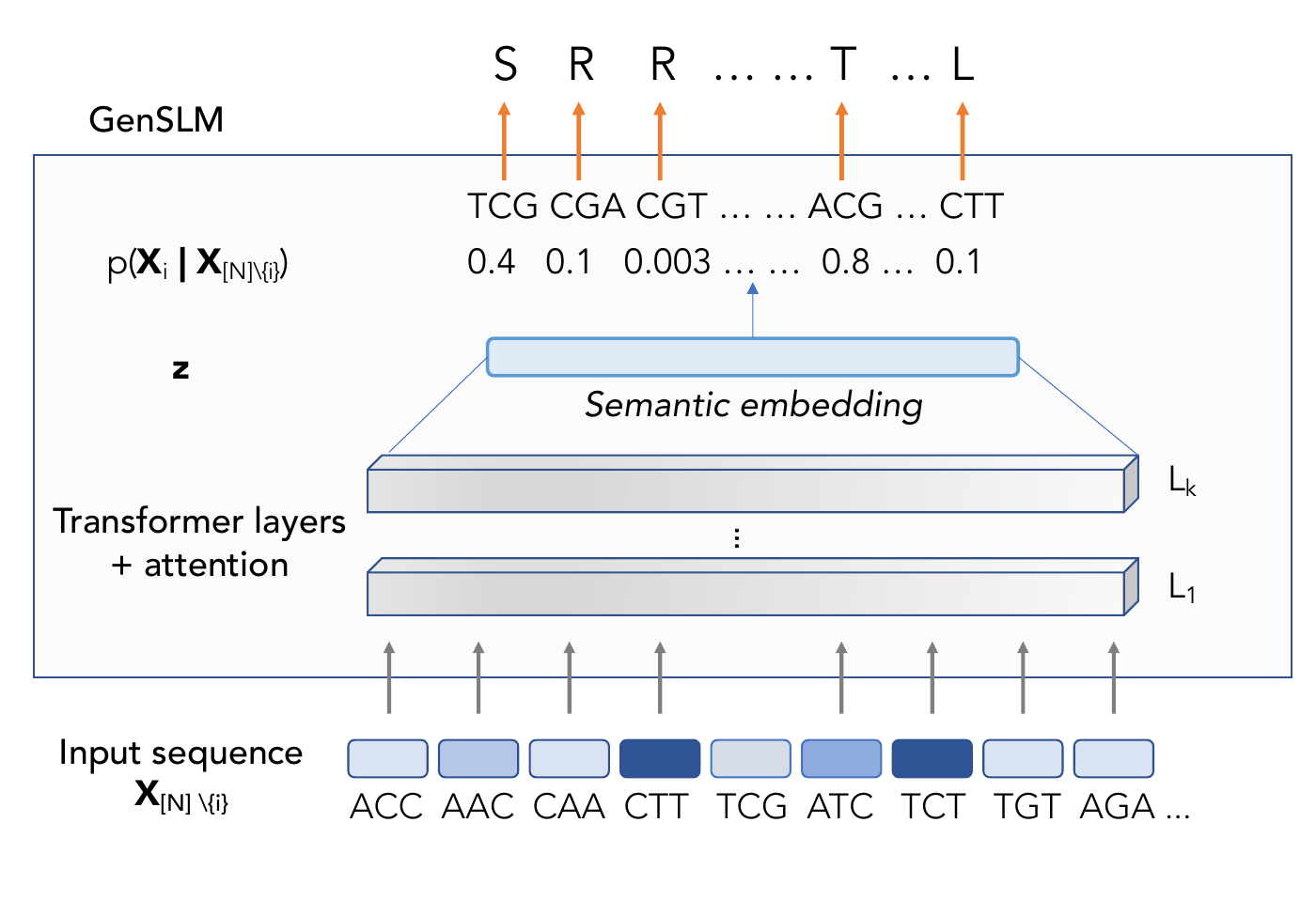} 
    \caption{Overview of GenSLM models for predictive modeling of SARS-CoV-2 evolution {\cite{genslm}}}
    \label{fig:LLM-genome}
     \vspace{-0.15in}
\end{figure}

(3) \textbf{GenSLM (Science Use Case)}: 
In addition to the benchmarks described above, we are also interested in evaluating how these models perform in real-world use cases. GenSLM \cite{genslm-source,genslm} is a genome-scale foundation model that can be generalized to other models. The goal of this model is to identify and classify different virus variants of concern, which can be then extended to gene or protein synthesis. It adapts GPT-based large language models with different parameters (25M-25B) with a 1.2B raw nucleotides dataset and is aimed at larger sequence lengths to help better capture the context and are generalizable to learn the evolution. 
Figure \ref{fig:LLM-genome} shows the overview of the GenSLM model that takes SARS-CoV-2 genomes (nucleotide sequences encoded at the codon level where every three nucleotides represent a codon) as input, which are fed to a series of transformer layers. The intermediate layers learn the semantic embedding of individual codons, which can be mapped to 29 individual virus-encoded protein sequences.
In this study, we focus on the evaluation of the GenSLM GPT model with 13B parameters on Nvidia A100, SambaNova SN30, and a 25B GPT3-XL model on Cerebras CS-2. 

\textbf{Metrics:} 
{Evaluating the performance of large language models goes beyond traditional metrics and includes considerations of computational efficiency and hardware utilization. As these models have grown in size and complexity, it is essential to assess their ability to process and generate text efficiently, especially considering the computational resources required for training and inference.}
{Throughput is a key performance metric that measures the rate at which a large language model can process a given amount of input data. It is often quantified in terms of tokens or sentences processed per second. Higher throughput values indicate better efficiency and the ability to handle large-scale language processing tasks effectively. In this work, we present the throughput in the number of tokens per second across the evaluated systems.}

{Hardware utilization is another important metric in evaluating large language models, as it assesses how effectively the computational resources are utilized during model training and inference. It involves multiple design choices, such as model parallelism, memory management, and efficient data processing techniques. Profiling the models to extract hardware utilization across the evaluated systems is ongoing and will be included in the final version.}

%% file: tex/05_Results.tex
\section{Implementation on AI Accelerators}
\label{section:accelerators}
The implementations of the evaluated models on each of the systems vary due to different software stacks and scales.
Here we describe the implementation details on each system for the three cases: transformer micro-benchmark, GPT-2 XL pre-training, and the GenSLM science application.

\subsection{Transformer micro-benchmark}
{The evaluation of the transformer micro-benchmark involved a meticulous implementation process aimed at assessing the computational efficiency and performance characteristics of this kernel on different AI accelerators. The transformer micro-benchmark is designed to simulate the core operation in the transformer models, which is widely utilized in various natural language processing tasks. The transformer micro-benchmark utilizes a layer in a standardized GPT-2 XL model with an input sequence of 1024, ensuring consistent and comparable results across different platforms. The implementation is carried out using the same deep learning framework, PyTorch, tailored to the unique capabilities of each platform. The workload is used to exploit parallelism and hardware-specific optimizations and to achieve optimal throughput and computational speed. Careful attention is paid to factors such as batch size to avoid bottlenecks and fully utilize the available hardware resources.}

{We use batch sizes of 8, 16, 32, and 64 for different configurations. Performance metrics, such as TFLOPS, are collected to quantify the capabilities of each hardware platform in handling the demanding computations required by the transformer models. This evaluation provides valuable insights into the strengths and weaknesses of different hardware when dealing with transformer-based workloads.}

\subsection{\label{subsec:GPT-2 XL-pretrain}GPT-2 XL pre-training}
%
As part of the GPT-2 XL study, we pre-train the model on the OWT dataset \cite{owt} where the raw data is pre-processed and tokenized for a given sequence length. The details of the tokenization and model implementation for each of the systems are described below.  
%

\textbf{\textit{Nvidia A100:}}{
%
We ran the models with varying micro-batch sizes, sequence lengths, and \textit{tensor-parallel} degree on different node counts, up to 64 A100 GPUs.
These are implemented with Megatron-DeepSpeed \cite{megatron-sc21} using \texttt{ZeRO Stage 1} with \texttt{fp16} precision. In these experiments, flash attention \texttt{flash-attn}~\cite{dao2022flashattention,dao2023flashattention2} was enabled, which is known to relieve memory pressure and generally improve throughput. 
BPE-based tokenizer with a vocabulary size of was used. The sequence lengths up to 2k were used. Experiments with Nvidia's NeMo framework for generative AI \cite{nvidia-nemo} are part of planned future work. 



\textbf{\textit{Cerebras CS-2:}} 
On the Cerebras CS-2 system, we ran the GPT-2 XL model implemented in the PyTorch framework with sequence lengths of 1024 and 2048 with a batch size of 112 on a single CS-2 engine. 
The implementation uses a custom GPT-2 tokenizer based on byte-level Byte-Pair-Encoding (BPE) with a vocab size of $50257$. The model is trained using mixed precision and precision opt level \cite{POL} of 1 (default) and 2. It uses an AdamW optimizer with a weight decay rate of 0.011. 
The model is trained with both dense and sparse configurations.
In the sparsity approach, all the weights of the dense layers are sparsified based on the degree provided. We ran the GPT-3 6.7B and GPT-3 30B models with various sparsity values. Here, a  sparsity value of $0.3$ means that 30\% of weights are pruned. The impact of sparsity on model throughput and loss are discussed in Section \ref{section:sparsity}. 

%
\textbf{\textit{SambaNova SN30:}} 
We evaluated pre-training performance on the OWT dataset on the SambaNova next-generation datascale SN30, which has 8 RDUs per node and 8 tiles per RDU.
We used the SN reference implementation of the GPT-1.5B model that fits on 4 tiles or half an RDU.
This implementation is based on the PyTorch-based SambaFlow framework and uses mixed precision (16 bit(i.e. bf16)  multipliers and 32-bit accumulators
It uses a BPE-based GPT-2 tokenizer with a vocab size of $50260$.  
We use data parallelism to scale across multiple tiles and nodes.
We scale for a maximum of 8 nodes (which corresponds to 128 instances of data-parallel runs) with each instance having a micro-batch size of 16. 
%

\textbf{\textit{Graphcore Bow Pod64}}: 
On the Bow Pod64, we leverage 64 IPUs to train the evaluated models.
The GPT-2 XL 1.5B model implemented in the PyTorch framework can be model-sharded across 4~IPUs. As part of the data pre-processing,  the implementation uses Nvidia's Megatron library for generating the training samples with a BPE tokenizer and a vocab size of $50272$.
The Poplar SDK uses the mapping on a Multiple Instruction Multiple Data (MIMD) fashion to leverage the IPUs and overlap computation and communication.
We used a local batch size of 1 in \texttt{FP16}, combined with large gradient accumulation step values of 128 and 1024.
Such large values help to minimize the communication overhead as it mimics a larger global batch size. 
We used a replication factor \cite{replication_factor}, of 1, 2, 4, and 16 to achieve better scaling, especially for smaller GAS values.
%

%
\textbf{\textit{Habana Gaudi2:}} 
We ran the GPT-2 XL model implemented in PyTorch with a sequence length of 1024 
with a local batch size of 32 on each HPU device.
The training of the GPT-2 XL model on Habana Gaudi2 represents a powerful and cutting-edge combination of software and hardware capabilities.
The data format used in training is \texttt{BF16}. The training samples are generated using a BPE tokenizer.


\textbf{\textit{AMD MI250:}} 
On the AMD MI250 system, we evaluated the performance of GPT-2  with absolute position embeddings trained with a causal language modeling (CLM) objective on the OWT dataset for up to 8 GPUs.
We leverage GPT-2 tokenizer based on byte-level Byte-Pair-Encoding. We used the reference PyTorch 2.0 implementation from Hugging Face to realize this. The performance is evaluated for sequence lengths of 1024 on batch sizes of 16 and 32 per GPU as well as GAS values of 1 and 4. 


\begin{figure*}[htpb]
     \begin{subfigure}[b]{0.33\textwidth}
 \includegraphics[width=\textwidth]{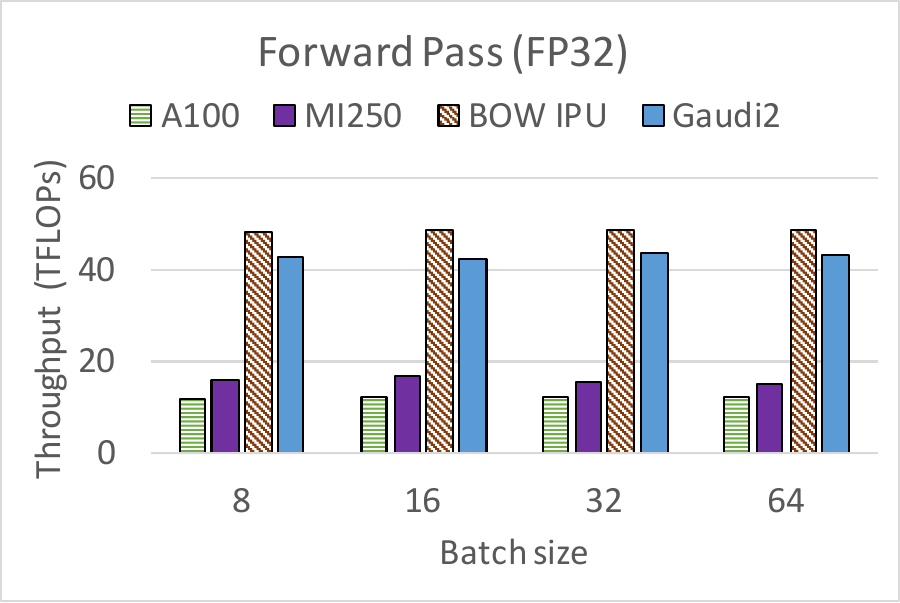}
  \caption{\texttt{FP32}}
     \end{subfigure}
     \begin{subfigure}[b]{0.33\textwidth}
        \includegraphics[width=\textwidth]{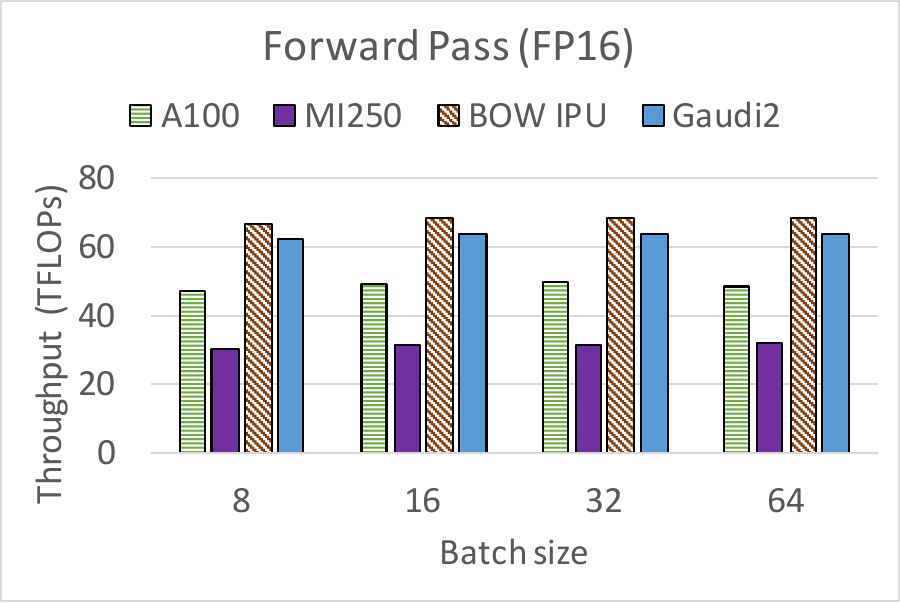}
         \caption{\texttt{FP16}}
     \end{subfigure}
     \hfill
     \begin{subfigure}[b]{0.33\textwidth}
        \includegraphics[width=\textwidth]{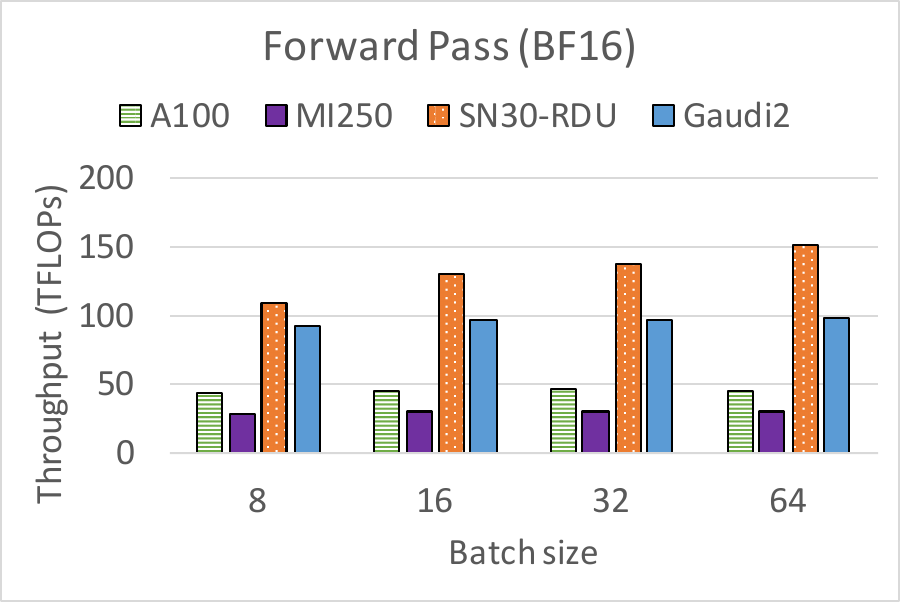}
         \caption{\texttt{BF16}}
     \end{subfigure}
     \caption{\label{fig:transformer-block-fwd}Throughput evaluation of transformer micro-benchmark in the forward pass with various precision}   \vspace{0.1in}
     \begin{subfigure}[b]{0.33\textwidth}
       \includegraphics[width=\textwidth]{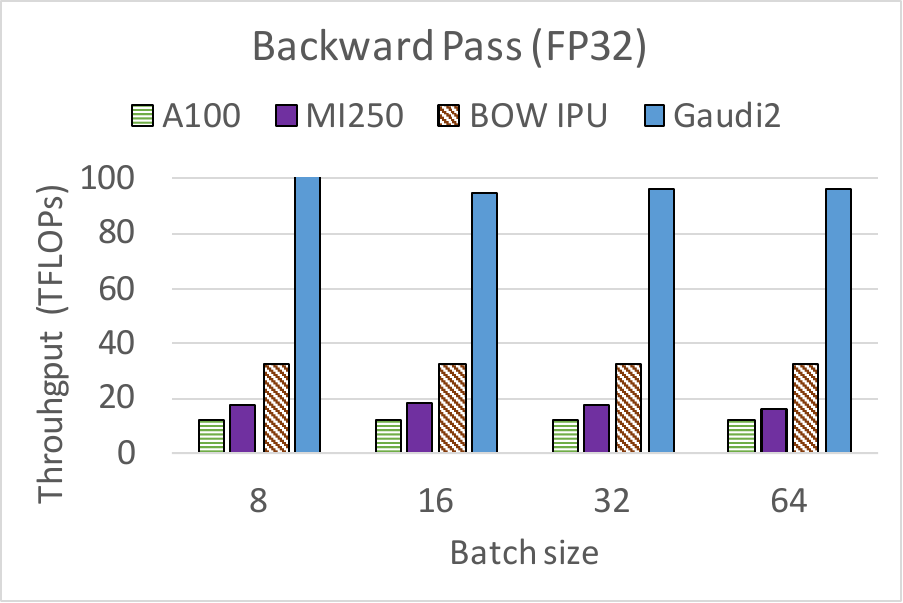}
       \caption{\texttt{FP32} }
     \end{subfigure}
     \begin{subfigure}[b]{0.33\textwidth}
        \includegraphics[width=\textwidth]{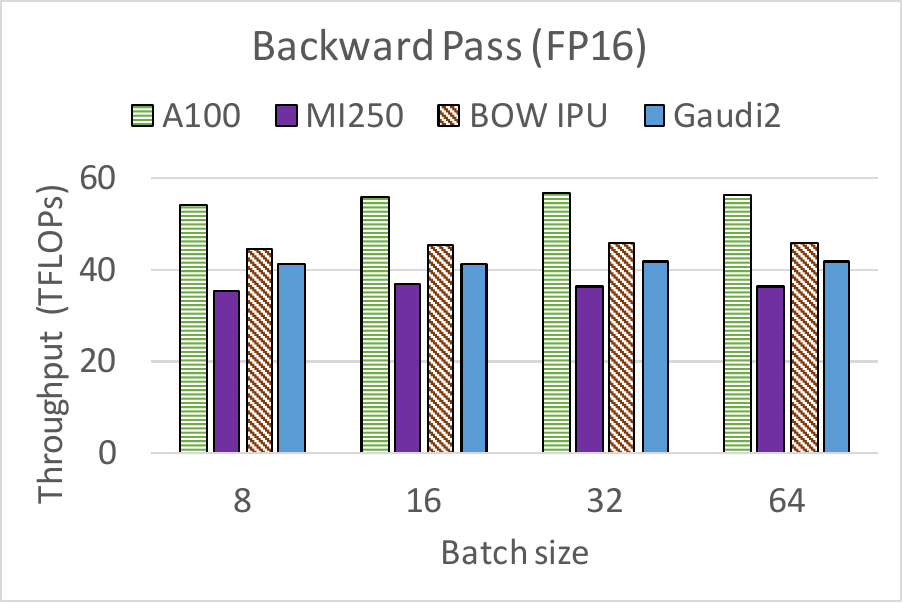}
         \caption{\texttt{FP16}}
     \end{subfigure}
     \hfill
     \begin{subfigure}[b]{0.33\textwidth}
        \includegraphics[width=\textwidth]{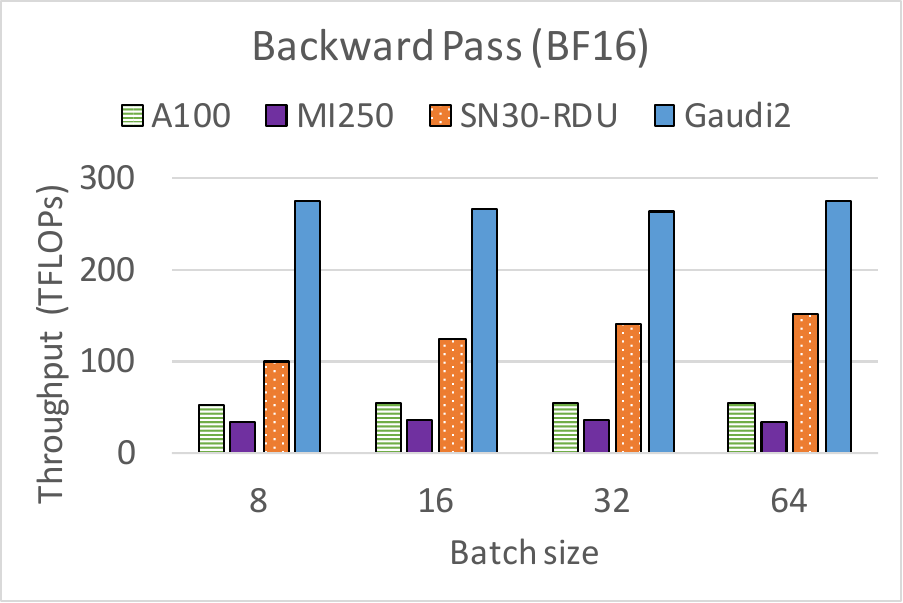}
         \caption{\texttt{BF16}}
     \end{subfigure}
     \caption{\label{fig:transformer-block-bkwd}Throughput evaluation of transformer micro-benchmark in the backward pass with various precision}
\end{figure*}

\subsection{GenSLM}
We implemented the GenSLM science application on Nvidia A100, SambaNova SN30, and Cerebras CS-2. The model is implemented using a PyTorch framework on all three systems. It uses the genomic
sequence dataset and tokenizes it using a codon-level
tokenizer that splits genomes into blocks of 3 nucleic acids. As the GenSLM application includes models with different number of model parameters ranging from 25M to 25B, 
we used two different model parameter sizes in this exercise. 
%
%

%
The SN30 implementation for GenSLM is based on the GPT-2 13B parameter model that uses a context length of 1024 with 44 layers, 64 attention heads, an embedding size of 4992, and a vocabulary size of 71. This GPT-2 13B parameter model with a batch size of 32 can be mapped within 4 tiles or equivalently half an RDU. 
The model used on Cerebras is a GPT-3 XL, a 1.5B parameter model that has 24 hidden layers and 16 attention heads. The model uses a vocabulary size of 70 and an embedding size of 2048. The model is trained for the genomic sequences of sequence length 10240 and a local size of 27. The other model parameters are similar to the GPT-2 XL implementation details as listed above. The GPT3-XL model is scaled across the two CS-2s to provide a global batch size of 54. 
On Nvidia A100, we use an identical GPT-2 13B model consisting of 40 layers of hidden dimension 5120 and 40 attention heads. 

\section{Results}
\label{section:results}
In this section, we present the experimental results for the three evaluation cases. We start with the transformer micro-benchmark evaluated with three precision in Section \ref{section:transformer}. Next, we present the results of the GPT-2 XL 1.5B parameter model, with a focus on scalability in Section \ref{section:scaling}, GAS study in Section \ref{section:gas}, sequence length analysis in Section \ref{section:seqlength}, and sparsity studies in Section \ref{section:sparsity}. Later, we detail our experimental results on the GenSLM model on three systems with models of three sizes: 1.5B, 13B, and 25B parameters in Section\ref{section:genslm}.


\subsection{Transformer micro-benchmark}
\label{section:transformer}
The results of the transformer micro-benchmark evaluation on a single NVIDIA A100 GPU, SambaNova SN30 RDU, Graphcore Bow IPU, Habana Gaudi2, and  AMD MI250\footnote{Evaluation of transformer micro-benchmark on CS-2 is under progress.} are shown in Figure~\ref{fig:transformer-block-fwd} and \ref{fig:transformer-block-bkwd}, which illustrate the throughput of forward and backward passes in three precision formats: \texttt{FP32, FP16, and BF16}.
Mi250 has a higher memory bandwidth (3276.8 GB/s) compared with that (2,039 GB/s) of A100, the higher bandwidth might be used to improve the performance of single precision. But the total efficiency of MI250 is lower than A100.


{It is observed that NVIDIA A100 GPU demonstrates the baseline throughput, capitalizing on its advanced tensor cores and parallel processing capabilities. The throughput with FP16 and BF16 precision is \~4x higher than FP32. A100 exhibits 2x theoretical performance for single precision compared to half precision. Additional performance improvement could arise due to a reduction in memory access using half-precision and this transformer is memory intensive kernel. 
SambaNova SN30, with its reconfigurable Dataflow architecture, exhibits impressive performance with BF16 precision, showcasing its potential for handling complex transformer workloads using the half-precision format.
Due to the pipelined/fusion execution on RDUs, there is naturally a warmup and cooldown phase as in any pipeline. More samples in a batch lead to longer steady-state behavior and higher effective throughput.
Graphcore Bow IPU, powered by IPUs, demonstrates exceptional performance with \texttt{FP32} and \texttt{FP16} precision, highlighting its suitability for NLP tasks.
Meanwhile, Habana Gaudi2 exhibits robust performance with all three formats, emphasizing its prowess in efficiently executing various transformer computations. In the backward pass, we think it’s due to higher utilization of hardware, and that brings higher throughput.
AMD MI250, capitalizing on its dedicated Tensor Processing Core array, exhibited remarkable acceleration and consistent throughput in the backward pass.}
%


\subsection{GPT-2 XL}%
For this model, we present throughput for the pre-training of the different configurations of the number of devices as a scaling study. Later we discuss the sensitivity of sequence length and gradient accumulation steps on the model throughput.
%
%
%

%
%
%
\subsubsection{Scaling study}
\label{section:scaling}
Here we present our findings from scaling the GPT-2 XL model on the different systems.
The number of devices used in the scaling study differs from one system to another due to the availability of resources.
In particular, we used 64 Nvidia A100 GPUs, 2 CS-2 engines, 64 SambaNova SN30 RDUs, 64 Graphcore Bow IPUs, 4 AMD MI250 GPUs, and 64 Habana Gaudi2 HPUs.
%
Figure \ref{fig:scalability}
shows the impact of model throughput (in log scale) across an increasing number of devices.
\footnote{On Bow, replica size 4 is not supported with 32 IPUs mesh topology due to an address limitation}. It is to be noted that the precision used on each system is different and the batch sizes are tuned for this configuration on each system.

The speedups across the number of devices on each evaluated system along with the scaling efficiencies are listed in Table \ref{tab:scaling-behavior}. A striking observation from this study showed that the model trained on 16 SN30 RDUs, 2 CS-2s, and 16 IPUs outperformed the runs on 64 A100s.\footnote{We aim to arrive at a fair comparison across the systems to the greatest possible extent.}
Additionally, Gaudi2 reported the highest scaling efficiency of 104\%. This scaling behavior is due to the optimizations from the Synapse software stack that help minimize the overhead in the sharding of the model and data across multiple HPUS. 
This is followed by Bow Pod64 which attains scaling efficiency of 100\%.
The superlinear scaling is enabled by the use of the replicated tensor sharding – as the scale is increased, the pressure on DRAM I/O is reduced for weight loading and IPU-Links are leveraged to exchange shards of weight tensors.
Additionally, Bow-IPU has 900 MB of SRAM and currently, it does not use DRAM for running this model hence, we would not be able to fit it into a single IPU due to the SRAM size limitation. We use 4 IPUs and run pipeline-parallel, with model layers distributed across the IPUs.
Cerebras CS-2's scales at 95.7\% efficiency, which demonstrates the efficiency of the weight streaming \cite{cs2-weightstreaming} technique with dedicated nodes, MemoryX for saving all model weights, and SwarmX to stream these weights to the compute engine. 
It is interesting to note that the SN30 and MI250 have scaling efficiencies around 80\% which is higher than the A100 number at 75.8\%. 

\begin{table}
    \centering
        \caption{Scaling behavior study with the GPT-2 XL model}
    \begin{tabular}{|p{1.2cm}|p{1cm}|p{1cm}|c|p{1.3cm}|} \hline 
         \textbf{System}&  \textbf{minimum \#devices}&  \textbf{maximum \#devices}& \textbf{scaling efficiency} & \textbf{speedup improvement} \\ \hline 
        Gaudi2&  1&  64& 104\% &  66.4x\\ \hline 
        Bow Pod64&  4&  64& 100.1\% &  16x\\ \hline  CS-2&  1&  2&  95.7\%& 1.8x\\ \hline 
         SN30&  1&  64& 82.6\%& 52.8x\\ \hline 
         MI250&  1&  4& 80\% &  3.2x\\ \hline
         A100&  4&  64& 75.8\% & 12.1x \\ \hline 
            
    \end{tabular}
    \label{tab:scaling-behavior}
\end{table}

 The results show that all the evaluated accelerators demonstrate increased throughput as the models are run across an increased number of devices. Though computationally expensive, models with a larger number of parameters can be trained better with more devices. As model sizes scale up in order of trillion parameters, it may not be feasible to run them on increased device count, thereby stressing the need to implement new approaches to fit larger models on fewer devices with the goal of improving the time to train a model given a computational budget.\footnote{Note that the \texttt{A100} and \texttt{IPU} results begin at 4 devices, and \texttt{CS-2} stops after 2.}


\begin{figure*}[htpb]
\includegraphics[width=\textwidth]{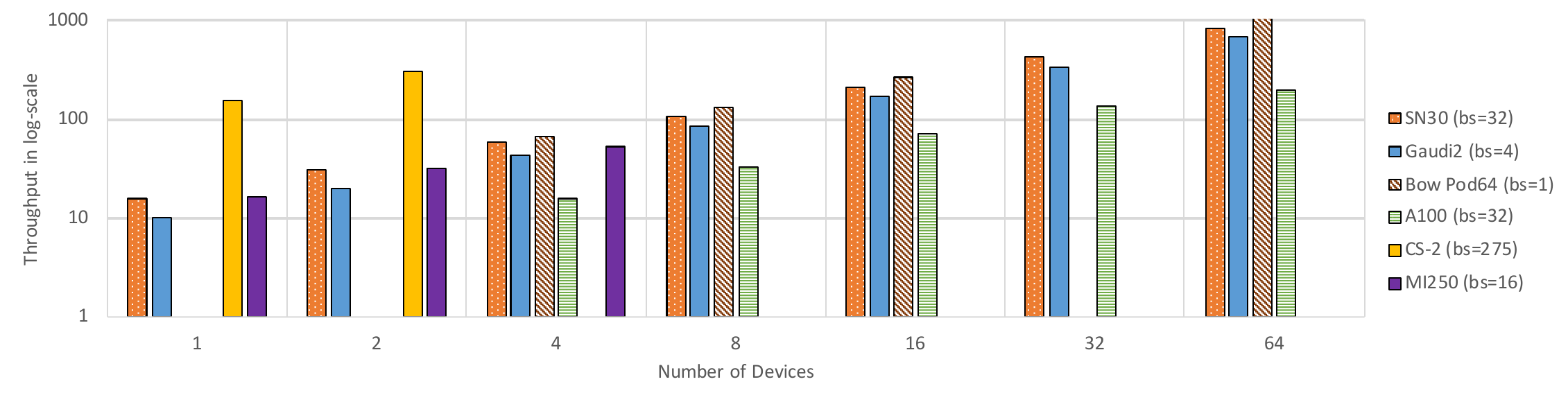}
\caption{GPT2-XL scaling study showing throughput in log scale with an increasing number of devices per accelerator with a sequence length 1K.}
 \label{fig:scalability}
\end{figure*}

\begin{figure*}[htpb]
     \centering
     \begin{subfigure}[t]{0.3\textwidth}
         \centering
        \includegraphics[width=\textwidth]{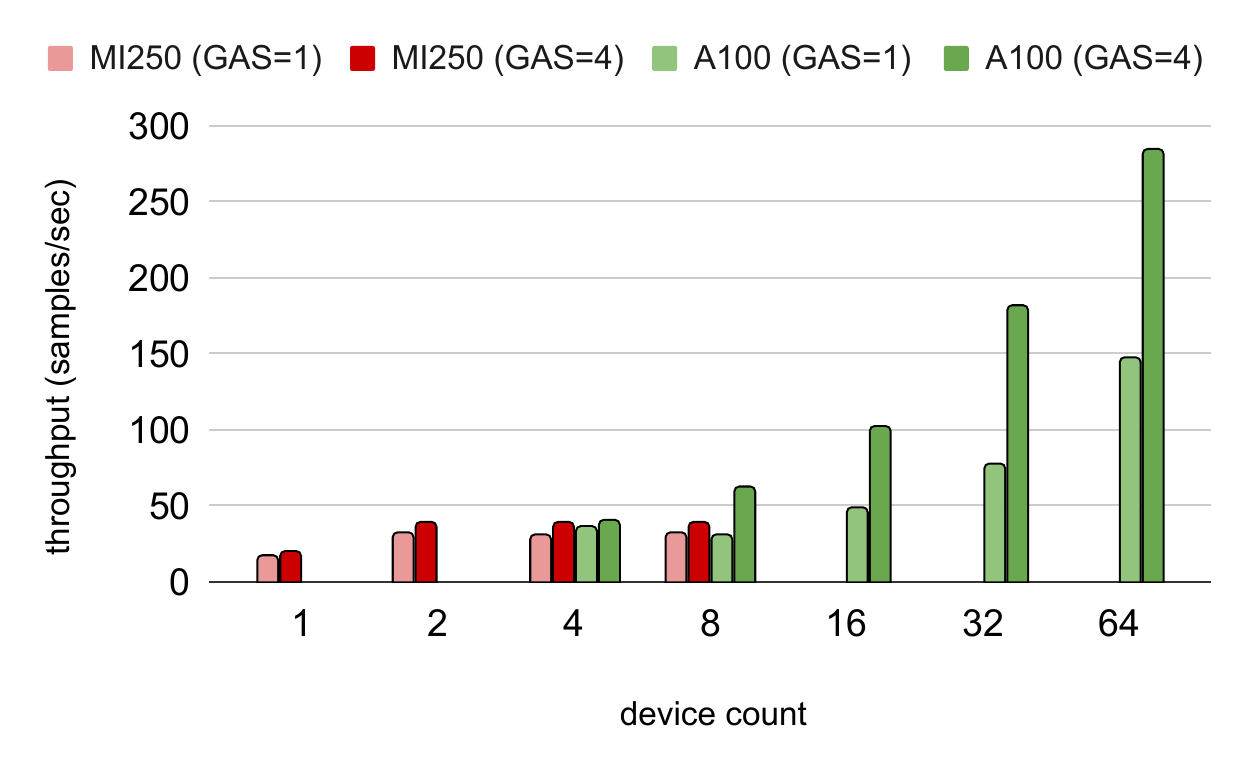}
         \caption{GAS study on Nvidia A100 \& AMD MI250}
     \end{subfigure}
      \begin{subfigure}[t]{0.3\textwidth}
         \centering
        \includegraphics[width=\textwidth]{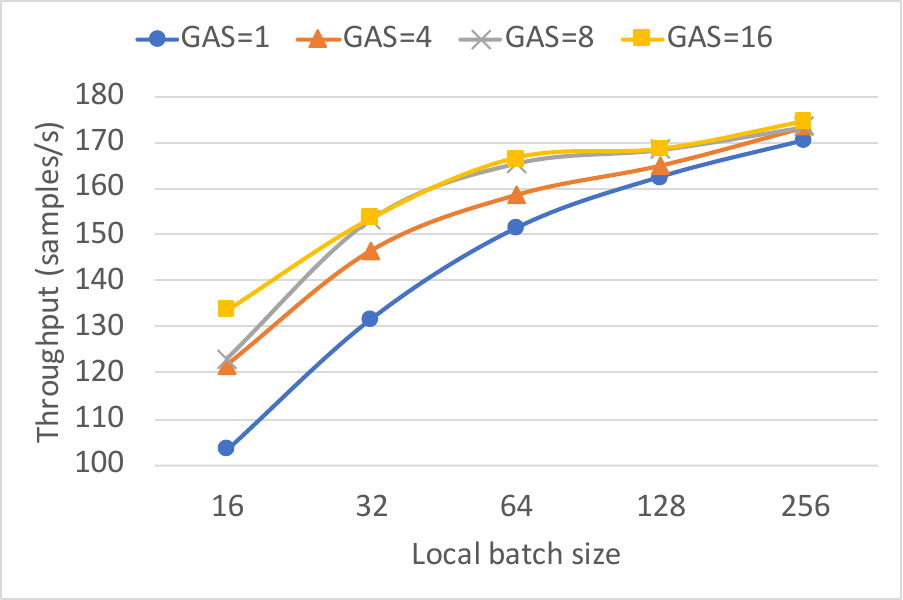}
         \caption{GAS study on SambaNova SN30}
     \end{subfigure}
        \begin{subfigure}[t]{0.3\textwidth}
         \centering
         \includegraphics[width=\textwidth]{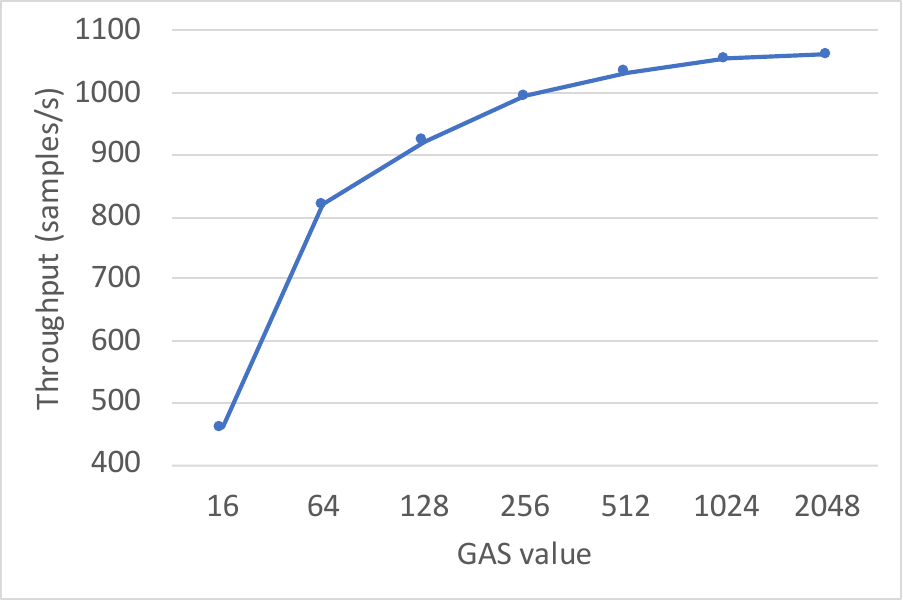}
         \caption{GAS study on Graphcore Bow Pod16 IPUs}
     \end{subfigure}
     \caption{Impact of Gradient Accumulation Step (GAS) value on model throughput }
        \label{fig:gas-study}
\end{figure*}

\subsubsection{Impact of sequence length}
\label{section:seqlength}
The need for scaling the sequence length in the large language model tasks has been highlighted in recent works~\cite{ding2023longnet} with scaling to more than 1B tokens. Large context lengths are extremely important in synthesizing genomic data as seen in the GenSLM science applications. Hence we study the impact of scaling the sequence lengths for the GPT-2 model and present the results on Nvidia A100, SambaNova SN30, and Cerebras CS-2 and exclude others either due to limited support or a work in progress.  
For this study, we use the GPT-2 XL 1.5B parameter model on both A100 and Cerebras system and both the Cerebras CS-2 and SambaNova SN30 use the GPT2-13B parameter model. As we can see from Table~\ref{table:SeqLength-throughput}, Nvidia A100 needs a minimum of 4 devices to fit a dataset of smaller sequence length. 
On the other hand, SambaNova SN30 and Cerebras CS-2, owing to their large local memory and compute architecture,  can fit models with longer sequence lengths on a single compute device. SambaNova SN30 can fit a 13B parameter model with varying sizes of sequence lengths ranging from 1K to 64K, all on a single RDU. The results are presented when we run 8 instances of the model on a node in data parallel mode. We see a predictable decline in the throughput with an increase in the sequence length. Moreover, from a sequence length of 32k, the implementation uses segmented softmax implementation.
We see a similar trend of sequence lengths impacting the throughput when the sequence length increases from 1K to 8K for the CS-2 system for a GPT-2 XL model. 

\begin{table}[htpb]
\small
\centering
    \caption{\label{table:SeqLength-throughput}Impact of Sequence length on model throughput}
    \begin{NiceTabular}{|p{1.75cm}|c|c|p{1.75cm}|}
    \toprule
    \textbf{System (model Size)}  & \textbf{Seq Length} & \textbf{Devices} & \textbf{Throughput (tokens/s)} \\
    \hline
    A100 (1.5B) & 1024 & 4  & 134,144 \\
                & 2048 & 4 & 124,928 \\
    \midrule
    CS-2 (1.5B) & 1024 & 1 & 133,069 \\
                & 2048 & 1 & 114,811 \\
                & 4096 & 1 & 63,488 \\
                & 8192 & 1 & 16,302 \\
    \midrule
    CS-2 (13B)  & 1024 & 1 & 20,685 \\
                & 2048 & 1 & 20,173 \\
                & 4096 & 1 & 17,531 \\
                & 8192 & 1 & 15,237 \\
                & 16384 & 1 & 11,796 \\
                & 32768 & 1 & 7537 \\
                & 51200 & 1 & 5120 \\
    \midrule
    SN30 (13B)  & 1024 & 8 & 22,135 \\
                & 2048 & 8 & 21,684 \\
                & 4096 & 8 & 17,000 \\
                & 8192 & 8 & 10,581 \\
                & 16384 & 8 & 4936 \\
                & 32768 & 8 & 5021 \\
                & 65536 & 8 & 1880 \\
    \toprule
    \end{NiceTabular}
\end{table}



\subsubsection{Impact of gradient accumulation steps}
\label{section:gas}
Large language models, especially those with billions of parameters, consume a significant amount of memory during training. Gradient accumulation allows for the accumulation of gradients over multiple mini-batches before updating the model's weights. This reduces the memory requirements compared to updating the model after processing each individual mini-batch. This is particularly important when working with hardware with limited memory capacity. In this study, we studied the impact of increasing GAS values on the model performance.

Figure \ref{fig:gas-study}(a) shows the sensitivity of the model throughput on GAS values on the A100 and MI250 GPUs. For this study, we test performance for GAS values of 1 and 4 while we keep batch size and sequence length constant at 16 and 1024 respectively. We observe throughput improvement for larger GAS values before it saturates as the number of devices increases.  
We see that when the GAS value is increased from 1 to 4 on the GPT-2 XL model with micro-batch size 1 and tensor parallel degree 1, the model throughput is increased by 1.74x across 64 A100 GPUs.

For MI250 GPUs, the model throughput is increased by 1.2x across 4 GPUs. The increased performance can be accounted for by the capacity to process more samples with increased GAS value. Also, for the A100 GPUs when we vary the GAS values from 1 through 32 for local batch size 32, we observed a 1.74x increase in the model throughput, thus confirming the source for this increase to larger GAS value at a relatively lower batch size. A detailed study on the trade-off between local batch size and GAS values for A100 and MI250s is a work in progress.
%
Figure \ref{fig:gas-study}(b) presents the throughput in samples per second for GAS values 1, 4, 8, and 16 on the SambaNova SN30 system. Though these results are provided for 16 instances of model training that trains across a full 8 RDU node, we can see similar behavior when extended to a full 8 racks. We observe that the throughput gain is significant with increasing GAS values when coupled with a smaller local batch size. On the contrary, increasing the GAS values has less or no effect on the throughput, which saturates at larger local batch sizes. This observation can be attributed to the fact that at larger batch sizes, the additional task of loading the gradient in case of the backward pass computations is significantly time-consuming compared to the time saved by reducing the number of optimization steps. 

Figure \ref{fig:gas-study}(c) shows the sensitivity of the model throughput on GAS values on the Graphcore POD16. For this study, we consider a replication factor of 4 indicating that a single instance of the model (that is sharded over 4 IPUs) is replicated 4 times to span an entire POD16 system. The results can be extended to a full rack of POD64. As we can see, Graphcore can support very large GAS values ranging from 16 to 2048. This is made possible by processing multiple batches and aggregating gradients into accumulator tensors without increasing memory usage. 
The SambaNova SN30 can also technically support significantly large GAS values, though its impact on the model throughput is yet to be investigated.
%
%
Cerebras gradient accumulation steps (GAS) are determined automatically by the Cerebras software stack to drive the best throughput given the requested global batch size and number of CS-2s. Given the limited user functionality to tune this parameter, we excluded CS-2 in this study.

\subsubsection{Sensitivity of weight sparsity on model throughput}
\label{section:sparsity}

Sparsity is important for training large language models primarily owing to several factors such as enhanced memory efficiency by storing only non-zero parameters, improved training speed due to a reduction of a number of parameters updated in each optimization step, and effective scalability. It also helps with faster inference as the computations are limited to non-zero weights.  
Here, we conducted a study of sensitivity of model sparsity on the throughput of the Cerebras CS-2.
The sparsity degree reflects the percentage of weights that are pruned, in order to accelerate the training.
From Figure \ref{fig:CS-2-sparsity} it can be observed that the model throughput increases from a dense model (s=0) to a highly sparse model (s=0.9).
For the GPT-2 XL model, we observed a throughput speedup of 1.8x to 2x with extreme sparsity (s=0.9) when compared to a completely dense model (s=0) \footnote{2-3x speedups with larger models}. Additionally, the sparsity degree has a higher impact on the throughput improvement with larger models. For the GPT-3 6.7B, a sparsity of 0.9 yields 2.1x over a dense model on 1 CS-2, and for GPT-3 30B, a sparsity of 0.9 yields 3.79x over a dense model on 1 CS-2. Further scaling out, this speedup factor is improved by 7.75x, and 15.49x for the 6.7B model on 8 and 16 CS-2s compared to 1 CS-2. 
%
This experiment demonstrated that model sparsity can significantly boost throughput and can potentially help to fit larger models on a relatively smaller number of devices.\footnote{The batch sizes used on CS-2s are the optimal ones for this model.}  

Additionally, Figure \ref{fig:CS-2-sparsity-losscurve} shows the loss curves for a GPT-3 model trained on 3B tokens, highlighting a 15\% loss in the model accuracy with sparsity s=0.9 as compared to a dense model.
During pre-training, there is a degradation in the training loss proportional to the sparsity value. 
%
%
However, dense finetuning of the model can recover such a gap, as past work by Cerebras on the sparse pre-training and dense fine-tuning framework has demonstrated. A 1.3B GPT-3XL model pre-trained with 75\% sparsity resulting in no significant loss in downstream task accuracy while reducing 2.5x pre-training FLOPs [citation]. We are currently exploring the sparse pretraining dense finetuning technique \cite{spdf} for large LLM models on CS-2.

Studies assessing the impact of sparsity on training on the SambaNova SN30 with techniques developed \cite{sambanova-sparsity} and Nvidia A100s are ongoing and will be included in the final version of the paper. 
Due to limited or no support of sparsity on the other systems, they were excluded from this study.

\begin{figure}[htpb]
\centering\includegraphics[width=\linewidth]{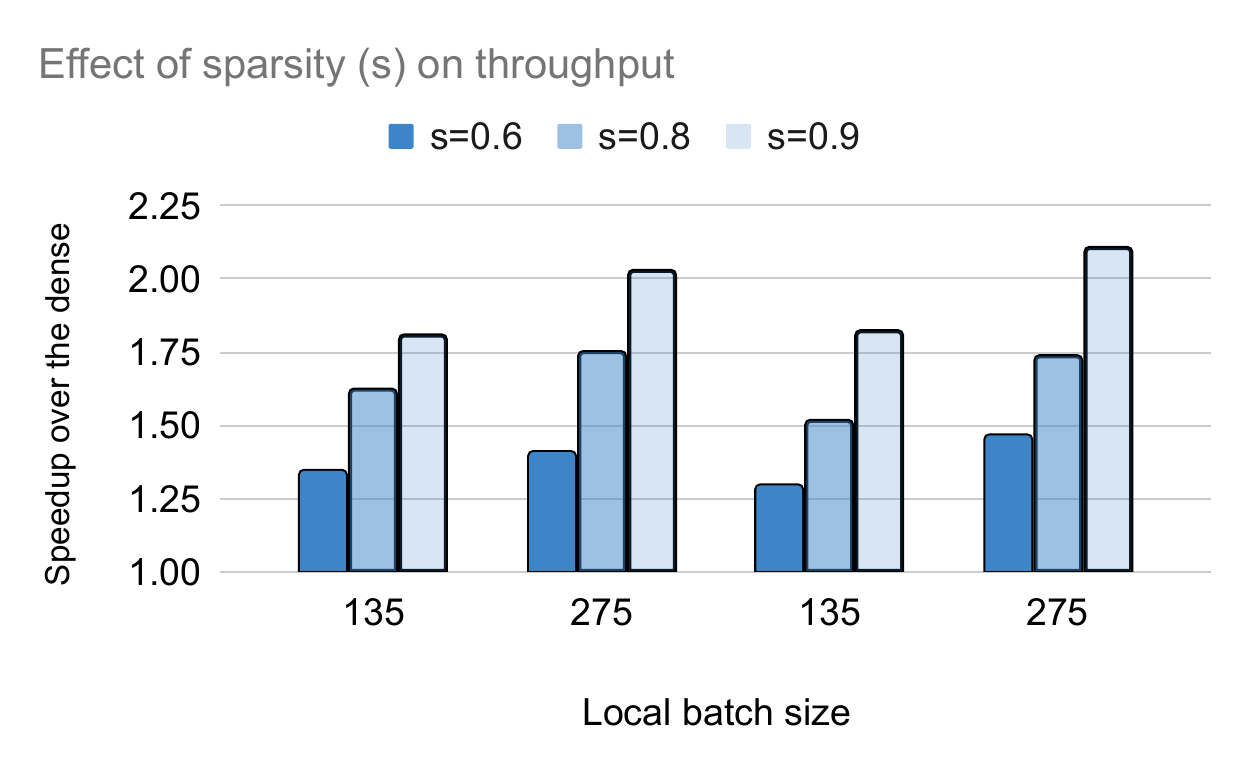}
 \caption{Effect of various sparsity levels (0, 0.6, 0.8, and 0.9) on model throughput on CS-2}
 \label{fig:CS-2-sparsity}
\end{figure}

\begin{figure}[htpb]
\centering
 \includegraphics[width=\linewidth]{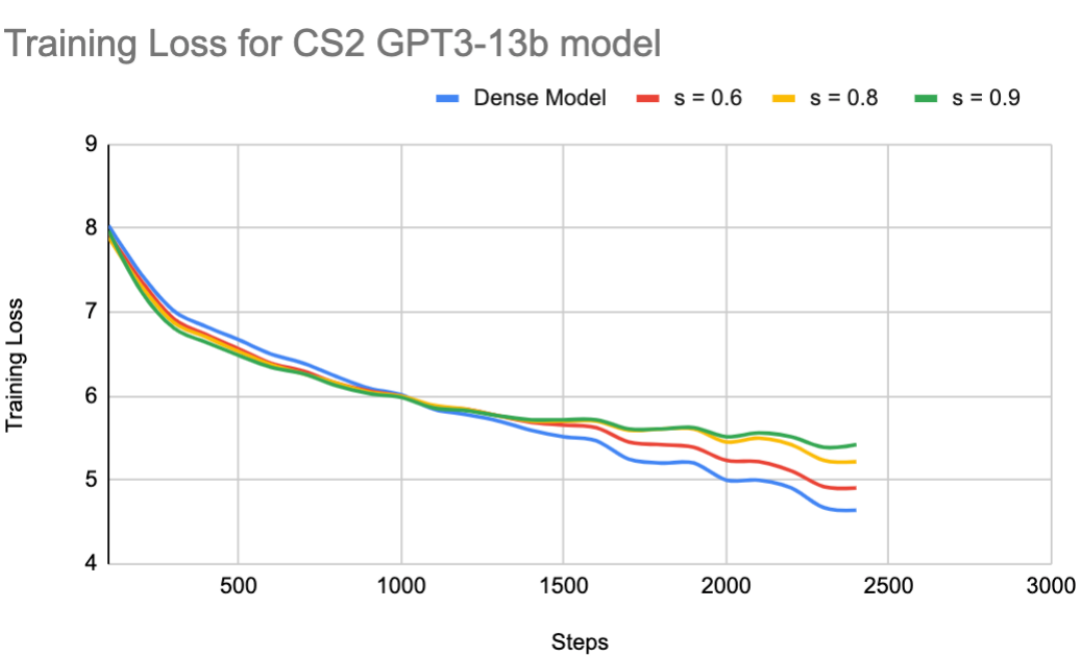}
 \caption{\label{fig:CS-2-sparsity-losscurve}Loss curve for sparse model implementation on CS-2}
\end{figure}

\begin{table}[htpb]
\centering
    \caption{\label{table:genslm-throughput}GenSLM model  performance evaluation }
    \begin{NiceTabular}{|p{2cm} | p{1cm} | p{1.5cm} | p{2cm} |}
    \toprule
    \textbf{System}  & \textbf{Devices} & \textbf{Sequence Length} & \textbf{Throughput (tokens/sec)} \\
    \hline
    A100 (1.5B) & 4 & 1024 & 41,891.84 \\
    CS-2 (1.5B) & 2 & 10240 & 173,875.2\\
    \hline
    A100 (13B) & 8 & 1024 & 1150.20 \\
    SN30 (13B) & 8 & 1024 & 6561.22 \\
    CS-2 (13B) & 1 & 1024 & 21,811.2 \\
    CS-2 (13B) & 1 & 10240 & 14,643.2 \\
    \midrule
     CS-2 (25B) & 1 & 10240 & 5324.8  \\
     \toprule
    \end{NiceTabular}
    \vspace{-5mm}
\end{table}

\subsection{GenSLM}
\label{section:genslm}

Table~\ref{table:genslm-throughput} presents the throughput of the GenSLM model, measured in tokens per second, across Nvidia A100, SambaNova SN30, and Cerebras CS-2. We present the throughput with three model sizes: A100 and CS-2 with a 1.5B parameter model, A100, SN30, and CS-2 with a 13B parameter model, and CS-2 with 25B parameter model.
In the case of A100, SN30, and CS-2, all implementations utilize the GPT-2 13B parameter model and are trained with a sequence length of 1024. Both A100 and the SN30 use a micro-batch size of 32, but since the SN30 can fit the 13B model of 32 micro-batch size on 4 tiles, it can accommodate 16 instances of the model in parallel over 8 devices\footnote{GenSLM runs on SN30 are not optimized at the time of writing this paper}. Different batch sizes were used for SN30 and A100 for the same model used in Table \ref{table:SeqLength-throughput}. 

Our observations reveal that when we consider the same number of compute units or devices, the SN30 demonstrates a remarkable 5.7x throughput improvement over the same number of A100s. We also find that a single Cerebras CS-2 demonstrates a massive 19x speedup for the same model when compared to eight A100s.

When we further compare the performance of the 1.5B parameter GenSLM model trained on Nvidia A100 and Cerebras CS-2, an interesting trend emerges. The CS-2 exhibits the capability to handle sequence lengths that are ten times longer than those run on the A100 GPUs, all while achieving a noteworthy 2x speedup improvement over the A100s.

This rigorous evaluation underscores the significant contribution of these accelerators in addressing large-scale real-world scientific applications. They play a pivotal role in accelerating the time to accuracy in the realm of large-scale scientific applications.

\subsection{Observations:}
This comprehensive benchmarking evaluation of one of the most pivotal AI workloads has yielded several noteworthy observations. Below, we present a few interesting and valuable insights.

\begin{itemize}
    \item The capacity to accommodate a sizable model within a single device is contingent upon the available computational resources and memory capacity. Even with the usage of powerful computational engines, the employment of innovative techniques aimed at minimizing memory consumption, particularly parameters such as weights and activations, is of paramount significance.

    \item Facilitating the execution of open-source models and streamlining the extension of models from Hugging Face are important to leverage these AI accelerators for a plethora of emerging AI for science applications.

    \item Achieving a fair comparison among AI accelerators presents notable challenges. Discrepancies arise from variations in local/micro-batch sizes, GAS values, the number of model replicas, and related factors. It is imperative to devise methodologies that facilitate a fair comparison.

    \item Notably, increasing GAS values does not invariably translate to  performance improvement beyond a certain threshold. This method, combined with a judicious choice of micro-batch size, enables for run at larger batch sizes.

    \item  Supporting longer sequence lengths is important to capture context, handle long-range dependencies, and excel in a wide range of tasks.

    \item With upcoming models with trillions of parameters and the requirement to cater to longer sequence lengths, 
 hardware and software design must be tailored to maximize computational capabilities while simultaneously minimizing memory usage.
\end{itemize}

%% file: tex/06_Conclusion.tex
\section{Conclusions}
\label{section:conclusion}

In this paper, we performed an extensive and comprehensive evaluation of generative AI models on non-traditional hardware with a focus on GPT models, an in-depth analysis of a core transformer block, and a science application (GenSLM). Additionally, we explored the scaling behavior of the GPT-2 model to gain insights into its performance characteristics. 
One of our important observations is the memory limitations inherent to the hardware that restrict the feasibility of the size of models that can fit on a single device. It further requires a distributed implementation, such as data parallel, with an increased number of devices. 
A near-linear scaling is also observed with an increased number of devices. Furthermore, the adoption of optimizations, such as weight sparsity, helps effectively reduce the communication overhead in distributed implementations.

We plan to continue this evaluation with a focus on longer sequence lengths and benchmark representative models for the emerging generative AI for science applications. We also would like to extend this comprehensive benchmarking study with larger models such as 175B parameter models and with generative models with distinct architectures such as Llama.
It has been observed that to facilitate effective training of significantly larger models, such as trillion parameters in AI for science applications, leveraging non-traditional AI hardware would be pivotal.
Optimizing other metrics besides model throughput, such as power consumption and I/O in training, particularly with increased computation and the utilization of larger data sets, will also be essential.

\section{Acknowledgment}
This work was supported by the Argonne Leadership Computing Facility, a U.S. Department of Energy (DOE) Office of Science User Facility, and by Laboratory Directed Research and Development (LDRD) funding from Argonne National Laboratory, provided by the Director, Office of Science, of the U.S. DOE under Contract No. DE-AC02-06CH11357.
We also extend our sincere thanks to the staff from Cerebras, SambaNova, Graphcore, and Habana who helped us to perform this study. 